\documentclass[conference,twocolumn,10pt]{IEEEtran}

\pdfoutput=1
\usepackage[cmex10]{amsmath}
\usepackage{amsfonts,amssymb}
\usepackage{verbatim}

\usepackage{cite}

\ifCLASSINFOpdf
  \usepackage[pdftex]{graphicx}
  \graphicspath{{graphics/}}
  \DeclareGraphicsExtensions{.pdf,.jpeg,.png}
\else
  \usepackage[dvips]{graphicx}
  \graphicspath{{graphics/}}
  \DeclareGraphicsExtensions{.eps}
\fi

\usepackage{array}

\usepackage{mdwmath}
\usepackage{mdwtab}
\DeclareMathOperator{\erf}{erf}

\begin{document}
\bibliographystyle{IEEEtran}
%
\title{Bounds on Distance Estimation via Diffusive Molecular Communication}

\author{\IEEEauthorblockN{Adam Noel$^{\ast}$, Karen C.
Cheung$^{\ast}$, and Robert Schober$^{\ast\dagger}$}
\IEEEauthorblockA{$^{\ast}$Department of Electrical and Computer
Engineering\\
University of British Columbia, Email: \{adamn, kcheung, rschober\}@ece.ubc.ca
\\ $^{\dagger}$Institute for Digital Communications\\
Friedrich-Alexander-Universit\"{a}t Erlangen-N\"{u}rnberg (FAU), Email:
schober@LNT.de}}


\newcommand{\dbydt}[1]{\frac{d#1}{dt}}
\newcommand{\pbypx}[2]{\frac{\partial #1}{\partial #2}}
\newcommand{\psbypxs}[2]{\frac{\partial^2 #1}{\partial {#2}^2}}
\newcommand{\dbydtc}[1]{\dbydt{\conc{#1}}}
\newcommand{\thev}{\theta_v}
\newcommand{\thevi}[1]{\theta_{v#1}}
\newcommand{\theh}{\theta_h}
\newcommand{\thehi}[1]{\theta_{h#1}}
\newcommand{\x}{x}
\newcommand{\y}{y}
\newcommand{\z}{z}
\newcommand{\rad}[1]{\vec{r}_{#1}}
\newcommand{\radmag}[1]{|\rad{#1}|}


\newcommand{\kth}[1]{k_{#1}}
\newcommand{\km}{K_M}
\newcommand{\vm}{v_{max}}
\newcommand{\conc}[1]{[#1]}
\newcommand{\conco}[1]{[#1]_0}
\newcommand{\C}{C}
\newcommand{\Cx}[1]{C_{#1}}
\newcommand{\CxFun}[3]{C_{#1}(#2,#3)}
\newcommand{\Cobs}{C_{obs}}
\newcommand{\Nobs}{{\Nx{\A}}_{obs}}
\newcommand{\Nobst}[1]{\Nobs\!\left(#1\right)}
\newcommand{\Nobsn}[1]{\Nobs\left[#1\right]}
\newcommand{\Nobsavgt}{\overline{{\Nx{\A}}_{obs}}(t)}
\newcommand{\Nobsavgtx}[1]{\overline{{\Nx{#1}}_{obs}}(t)}
\newcommand{\Nobsavg}[1]{\overline{{\Nx{\A}}_{obs}}\left(#1\right)}
\newcommand{\Nobsavgmax}{\overline{{\Nx{\A}}_{max}}}
\newcommand{\Nnoisetavg}[1]{\overline{{\Nx{\A}}_{n}}\left(#1\right)}
\newcommand{\Nxavg}[1]{\overline{{\Nx{\A}}_{#1}}}
\newcommand{\Nxtavg}[2]{\overline{{\Nx{\A}}_{#2}}\left(#1\right)}
\newcommand{\Nxt}[2]{{\Nx{\A}}_{#2}^\star\left(#1\right)}
\newcommand{\DMLSNntavg}[1]{\overline{{\Nx{\DMLSA}}_{n}^\star}\left(#1\right)}
\newcommand{\DMLSNnavg}{\overline{{\Nx{\DMLSA}}_{n}^\star}}
\newcommand{\DMLSNxavg}[1]{\overline{{\Nx{\DMLSA}}_{#1}^\star}}
\newcommand{\DMLSNxtavg}[2]{\overline{{\Nx{\DMLSA}}_{#2}^\star}\left(#1\right)}
\newcommand{\DMLSNxt}[2]{{\Nx{\DMLSA}}_{#2}^\star\left(#1\right)}
\newcommand{\DMLSNx}[1]{{\Nx{\DMLSA}}_{#1}^\star}
\newcommand{\Nnoiset}[1]{{\Nx{\A}}_{n}\left(#1\right)}
\newcommand{\Ntxt}[1]{{\Nx{\A}}_{TX}\left(#1\right)}
\newcommand{\Ntxtavg}[1]{\overline{{\Nx{\A}}_{TX}}\left(#1\right)}
\newcommand{\DMLSNtxtavg}[1]{\overline{{\Nx{\DMLSA}}_{tx}^\star}\left(#1\right)}
\newcommand{\DMLSNtxt}[1]{{\Nx{\DMLSA}_{tx}^\star}\left(#1\right)}
\newcommand{\DMLSNtxtavgU}[2]{\overline{{\Nx{\DMLSA}}_{tx,#2}^\star}\left(#1\right)}
\newcommand{\DMLSNtxtU}[2]{{\Nx{\DMLSA}_{tx,#2}^\star}\left(#1\right)}
\newcommand{\Cgen}{C_A(r, t)}
\newcommand{\radbind}{r_B}

\newcommand{\M}{M}
\newcommand{\smM}{m}
\newcommand{\A}{A}
\newcommand{\B}{B}
\newcommand{\X}{S}
\newcommand{\vx}[1]{v_{#1}}
\newcommand{\vpara}{v_{\scriptscriptstyle\parallel}}
\newcommand{\vperp}{v_{\perp}}
\newcommand{\vxvec}[1]{\vec{v}_{#1}}
\newcommand{\Pec}[1]{v^\star_{#1}}
\newcommand{\Pecper}{\Pec{\perp}}
\newcommand{\Pecpara}{\Pec{\scriptscriptstyle\parallel}}
\newcommand{\metre}{\textnormal{m}}
\newcommand{\second}{\textnormal{s}}
\newcommand{\molecule}{\textnormal{molecule}}
\newcommand{\bound}{\textnormal{bound}}
\newcommand{\argmax}{\operatornamewithlimits{argmax}}
\newcommand{\Dx}[1]{D_{#1}}
\newcommand{\Nx}[1]{N_{#1}}
\newcommand{\Nemit}{\Nx{EM}}
\newcommand{\Da}{D_\A}
\newcommand{\En}{E}
\newcommand{\en}{e}
\newcommand{\Ne}{\Nx{\En}}
\newcommand{\De}{D_\En}
\newcommand{\EA}{EA}
\newcommand{\ea}{ea}
\newcommand{\Nint}{\Nx{\EA}}
\newcommand{\Di}{D_{\EA}}
\newcommand{\Etot}{\En_{Tot}}
\newcommand{\stepl}{r_{rms}}
\newcommand{\AP}{A_P}
\newcommand{\Ri}[1]{R_{#1}}
\newcommand{\ro}{r_0}
\newcommand{\rone}{r_1}
\newcommand{\visc}{\eta}
\newcommand{\bolt}{\kth{B}}
\newcommand{\temp}{T}
\newcommand{\T}{T_{int}}
\newcommand{\Vobs}{V_{obs}}
\newcommand{\robs}{r_{obs}}
\newcommand{\Ve}{V_{enz}}
\newcommand{\tint}{\delta t}
\newcommand{\tmax}{t_{max}}
\newcommand{\Cobsfrac}{\alpha}
\newcommand{\dist}{L}
\newcommand{\DMLSA}{a}
\newcommand{\DMLSt}[1]{t_{#1}^\star}
\newcommand{\DMLSx}{x^\star}
\newcommand{\DMLSy}{y^\star}
\newcommand{\DMLSz}{z^\star}
\newcommand{\DMLSr}[1]{r_{#1}^\star}
\newcommand{\DMLSrad}[1]{\rad{#1}^\star}
\newcommand{\DMLSradmag}[1]{|\DMLSrad{#1}|}
\newcommand{\DMLSC}[1]{\Cx{#1}^\star}
\newcommand{\DMLSCxFun}[3]{{\DMLSC{#1}}(#2,#3)}
\newcommand{\DMLSc}[1]{\gamma_{#1}}
\newcommand{\DMLSV}{\Vobs^\star}
\newcommand{\DMLSNA}{\overline{{\Nx{\DMLSA}}_{obs}^\star}(\DMLSt{})}
\newcommand{\DMLSNtx}{\overline{{\Nx{\DMLSA}}_{TX}^\star}(\DMLSt{})}
\newcommand{\DMLSNAb}{\overline{{\Nx{\DMLSA}}_{obs}^\star}(\DMLSt{B})}
\newcommand{\DMLSNAmax}{{\overline{{\Nx{\DMLSA}}_{max}^\star}}}
\newcommand{\DMLStmax}[1]{{\DMLSt{#1}}_{,max}}
\newcommand{\DMLSdim}{\mathcal{D}}
\newcommand{\DMLSthreshInt}{\alpha^\star}
\newcommand{\DMLSv}[1]{v^\star_{#1}}
\newcommand{\DMLSvxvec}[1]{\vec{v}^\star_{#1}}

\newcommand{\data}[1]{W\left[#1\right]}
\newcommand{\dataSeq}{\mathbf{W}}
\newcommand{\dataSet}{\mathcal{W}}
\newcommand{\dataObs}[1]{\hat{W}\left[#1\right]}
\newcommand{\numX}[2]{n_{#1}\left(#2\right)}
\newcommand{\thresh}{\xi}
\newcommand{\poissBar}{\Big|_\textnormal{Poiss}}
\newcommand{\gaussBar}{\Big|_\textnormal{Gauss}}
\newcommand{\textBar}[1]{\Big|_\textnormal{#1}}
\newcommand{\eqBar}[2]{\Big|_{#1 = #2}}
\newcommand{\Pobs}{P_{obs}}
\newcommand{\Pobsx}[1]{P_{obs}\left(#1\right)}
\newcommand{\Pone}{P_1}
\newcommand{\Pzero}{P_0}
\newcommand{\Pe}[1]{P_{err}\left[#1\right]}
\newcommand{\Peavg}[1]{\overline{P_{err}}\left[#1\right]}
\newcommand{\threshInterval}{\alpha}
\newcommand{\pstay}[1]{P_{stay}\left(#1\right)}
\newcommand{\pleave}[1]{P_{leave}\left(#1\right)}
\newcommand{\parrive}[1]{P_{arr}\left(#1\right)}

\newcommand{\VAmem}{F}
\newcommand{\VAstate}{f}
\newcommand{\VAdataObs}[2]{\hat{W}_{\VAstate_{#2}}\left[#1\right]}
\newcommand{\VAcurLL}[2]{\Phi_{\VAstate_{#2}}\left[#1\right]}
\newcommand{\VAcumLL}[2]{L_{\VAstate_{#2}}\left[#1\right]}

\newcommand{\weight}[1]{w_{#1}}

\newcommand{\fof}[1]{f\left(#1\right)}
\newcommand{\floor}[1]{\lfloor#1\rfloor}
\newcommand{\lam}[1]{W\left(#1\right)}
\newcommand{\EXP}[1]{\exp\left(#1\right)}
\newcommand{\ERF}[1]{\erf\left(#1\right)}
\newcommand{\SIN}[1]{\sin\left(#1\right)}
\newcommand{\SINH}[1]{\sinh\left(#1\right)}
\newcommand{\COS}[1]{\cos\left(#1\right)}
\newcommand{\COSH}[1]{\cosh\left(#1\right)}
\newcommand{\Ix}[2]{I_{#1}\!\left(#2\right)}
\newcommand{\Jx}[2]{J_{#1}\!\left(#2\right)}
\newcommand{\E}[1]{E\left[#1\right]}
\newcommand{\GamFcn}[1]{\Gamma\!\left(#1\right)}
\newcommand{\mean}[1]{\mu_{#1}}
\newcommand{\var}[1]{\textnormal{var}\!\left(#1\right)}
\newcommand{\mse}[1]{\textnormal{mse}\!\left(#1\right)}

\newcommand{\w}{w}
\newcommand{\n}{n}
\newcommand{\gx}[1]{g\left(#1\right)}
\newcommand{\hx}[1]{h\left(#1\right)}
\newcommand{\tx}[1]{t\left(#1\right)}
\newcommand{\ux}[1]{u\left(#1\right)}
\newcommand{\deltObs}{t_{o}}
\newcommand{\sx}[1]{s_{#1}}
\newcommand{\constAx}[1]{\Lambda_{#1}}
\newcommand{\constBx}[1]{\Phi_{#1}}
\newcommand{\constCx}[1]{\Psi_{#1}}

\newcommand{\new}[1]{\textbf{#1}}
\newcommand{\ISI}{ISI}
\newcommand{\DDFSE}{DDFSE}
\newcommand{\PDF}{PDF}
\newcommand{\CDF}{CDF}
\newcommand{\AWGN}{AWGN}

\newtheorem{theorem}{Theorem}

\maketitle

\begin{abstract}
This paper studies distance estimation for diffusive molecular
communication. The strength of the channel impulse response generally decreases with distance, so it is measured to estimate the distance.
The Cramer-Rao
lower bound on the variance of the distance estimation error
is derived. The lower bound is derived for a physically unbounded environment
with molecule degradation and steady uniform flow.
The maximum likelihood distance estimator is derived and its accuracy
is shown via simulation to perform very close to the Cramer-Rao lower
bound. An existing protocol is shown to be equivalent to the maximum likelihood
distance estimator if only one observation is made. Simulation results also show
the accuracy of existing protocols with respect to the Cramer-Rao lower bound.
\end{abstract}

\section{Introduction}

Applications in areas such as biological engineering and manufacturing
could potentially be improved by molecular communication (MC); see
\cite{RefWorks:801}.
MC is characterized by the encoding of information into
molecules that are released by transmitters and are then transported to their
intended destinations. Communication via diffusion, where the
molecules that are released move randomly due to collisions with other molecules
in the environment, is particularly advantageous for ad hoc networks of small
devices because there are no fixed
connections between devices and no external energy is required for molecule propagation.

Even though diffusive communication is commonly used in biological systems where
small molecules need to quickly travel short distances (see \cite[Ch.
4]{RefWorks:588}), the diffusive transmission of arbitrarily large amounts of
information, as would be desired for the aforementioned
applications, is severely limited by intersymbol interference (ISI). Signal
processing techniques have been proposed to mitigate the impact of ISI, such as
those described in \cite{RefWorks:644,RefWorks:747,RefWorks:786}.
However, the implementation of such techniques relies critically on the
knowledge of the expected channel impulse response, i.e., the number of
molecules \emph{expected} at a receiver over time given that molecules were
instantaneously released by a transmitter.

The expected channel impulse response
is generally a function of the parameters of the physical environment,
including the distance between the devices, the diffusion coefficient,
whether there is any flow, and whether the information molecules can participate
in chemical reactions. Any of these parameters might also change over time.
Depending on the application, specific parameters might need to be individually
estimated. For example, knowledge of the distance between devices is
essential for applications such as tuning the spatial
distribution of a network and
device addressing via nodes that release
molecules continuously; see \cite{RefWorks:608,RefWorks:488}.

We are interested in studying the problem of distance
estimation for diffusive molecular communication.
The number of molecules observed from an impulsive release decreases with distance. Thus, we can use the number of molecules observed to estimate the distance.
We wish to obtain bounds on estimation so that we can assess the
accuracy of any estimator relative to the bound. Existing
work has already introduced protocols that we can compare with a bound on
distance estimation; see
\cite{RefWorks:802,RefWorks:614,RefWorks:776,RefWorks:674}.

Protocols for measuring distance were first described for 1-dimensional
environments in \cite{RefWorks:802}, where the authors introduced what we call
feedback or \emph{two-way} methods that relied on either an instantaneous (i.e.,
signal spike or impulse) or a continuous release of molecules. We label these
methods two-way because two devices release molecules for one
of them to estimate the distance.
The same authors expanded their study of
impulse-based protocols in \cite{RefWorks:614}, where they used the standard
deviation of the estimation error as a metric to evaluate and compare protocol
performance. In \cite{RefWorks:776}, impulse-based distance estimation protocols
were introduced for 1-dimensional environments such that no feedback signal must be
sent, i.e., \emph{one-way} protocols. These protocols use
additional signal processing at the device where the estimate is made,
and use data from multiple observations. Furthermore, the authors
proposed a method for measuring the distance by releasing \emph{multiple}
impulses. A 2-dimensional environment was studied in \cite{RefWorks:674}, where
two-way continuous release and impulse methods were compared, and molecules
were either observed or captured until a threshold value was reached.

All existing distance estimation protocols are heuristic, i.e., they were
designed based on knowledge of the expected channel impulse response but not
with respect to an optimality criterion. The performance of any given protocol
has been evaluated in comparison with other protocols and the knowledge of the
true distance between devices.
However, existing work has not offered insight into the optimality of any
protocol.

Suppose that we derive the best possible performance of \emph{any} estimator
under some criterion, given the knowledge required for its implementation.
If we know the best possible estimator performance,
then we can perform a
complete assessment of protocols because we
can evaluate whether a given protocol is making the most accurate estimate for
the knowledge that it requires.
For example, we would expect that a simple protocol estimating the distance
using a single observation would be less accurate than a protocol that makes
multiple observations over time. But, the accuracy of the simpler protocol
might be much closer to the corresponding optimal accuracy and might provide
insight into improving the performance of more complex estimators. Furthermore,
if we know that a given estimator is optimal, then we know that we cannot
improve the estimator without incorporating additional information such as more
observations.

In this paper, we derive a bound on the accuracy of
distance estimation protocols as a function of the observations of molecules that are made to calculate the estimate. The primary contributions of this paper are
summarized as follows:
\begin{enumerate}
    \item We derive the Cramer-Rao lower bound (CRLB) on the variance of the
    estimation error of any unbiased distance estimator as a function of the number of independent observations of a transmitted impulse. We
    derive the CRLB for an unbounded 3-dimensional fluid environment with
    steady uniform flow and first-order molecule degradation.
    \item We derive the maximum likelihood distance estimator for any number of
    independent observations of an impulse signal. A closed-form solution exists
    if there is only one observation, and we show that this special case is
    equivalent to an existing protocol. In the case of
    multiple observations, we perform a numerical
    search to find the maximum likelihood estimate.
    \item We extend a selection of existing distance estimation protocols to the
    physical environment considered in this paper. We transform two-way
    protocols into one-way form in order to directly compare all protocols.
\end{enumerate}
    
The rest of this paper is organized as follows.
The physical environment is described and existing protocols for distance
estimation are reviewed in Section~\ref{sec_model}.
In Section~\ref{sec_bound}, we derive the CRLB
on the variance of the estimation error.
In Section~\ref{sec_est}, we derive the maximum likelihood distance estimator.
Numerical and simulation results
are presented in Section~\ref{sec_results}. Conclusions are drawn
in Section~\ref{sec_concl}.

\section{System Model and Preliminaries}
\label{sec_model}

In this section, we describe the diffusive communication environment,
including the two devices between which the distance must be estimated.
We present the channel impulse response and derive properties that will
be useful when designing and analyzing distance estimation protocols.
We also review existing protocols for distance estimation and describe how
they are executed as one-way methods.

\subsection{Physical Environment}

We consider an unbounded, 3-dimensional fluid environment with uniform constant
temperature and viscosity. There are two fixed devices, which
we label the transmitter (TX) and the receiver (RX) because we are considering
\emph{one-way} distance estimation. The RX is a sphere of radius $r_{RX}$ and
volume $V_{RX}$. The coordinate axes are defined by
placing the RX at the origin and the TX at Cartesian
coordinates $\{-d,0,0\}$, such that $d$ is the distance that is
to be measured. There is a steady uniform flow $\vxvec{}$ defined by two
velocity components, i.e., the component in the direction of a line pointing
from the TX to the RX, $\vpara$, and the component perpendicular
to this line, $\vperp$. We note that $\vperp$ is the component of $\vxvec{}$
that lies along the $yz$-plane but, due to symmetry, the
precise direction of this component is not required.

The TX releases type $\A$ molecules, which can be detected by
the RX. We assume that the RX knows the time when $\A$ molecules are
released\footnote{Synchronization strategies include blind
synchronization via a maximum likelihood approach in \cite{RefWorks:761} and
using inhibitory feedback molecules in \cite{RefWorks:773}.}.
The constant diffusion coefficient of
the $\A$ molecules is $\Dx{}$. The $\A$ molecules can also degrade
into a form that is not recognizable by the RX via a first-order
chemical reaction that can be described as
\begin{align}
\label{k_mechanism}
& \A \xrightarrow{\kth{}} \emptyset,
\end{align}
where $\kth{}$ is the first-order reaction rate
constant in $\second^{-1}$.
We ignore the chemical kinetics of the reception process at the RX by modeling the RX as a passive observer
that does not impede the diffusion of the $\A$ molecules. By ignoring the reception kinetics, we facilitate our analysis and emphasize the impact of
the propagation environment.
As an observer, the RX can perfectly
count the number of $\A$ molecules within its volume $V_{RX}$ at any time
instant.

\subsection{Analytical Preliminaries}

We require the expected channel impulse response at the RX
for the design and analysis of distance estimation protocols.
The expected channel impulse response is the number of molecules
expected at the RX due to an instantaneous release of
molecules by the TX. In this paper, we apply the uniform concentration
assumption, i.e., we assume that the $\A$ molecule concentration expected at the
RX due to molecules released by the TX is uniform
throughout the RX and equal to that expected at the center of the
RX. We have previously studied the accuracy of this assumption in
flowing environments in \cite{RefWorks:752}. If $\Nx{\A_{EM}}$ molecules
are instantaneously released by the TX at time $t=0$, then the number of
molecules expected to be observed by the RX, $\Nobsavgtx{\A}$, is given by
\cite[Eq. (12)]{RefWorks:747}
\begin{equation}
\label{trans1_impulse}
\Nobsavgtx{\A} = \frac{\Nx{A_{EM}}V_{RX}}{(4\pi \Dx{}
t)^{3/2}}\EXP{-\kth{}t - \frac{\radmag{eff}^2}{4\Dx{} t}},
\end{equation}
where $\radmag{eff}^2 = (d - \vpara t)^2 + (\vperp t)^2$ is the
square of the \emph{effective} distance from the TX to the RX.

All existing impulse-based distance estimation protocols rely on re-arranging
(\ref{trans1_impulse}), where all parameters but $d$ are either given,
observed, or removed via substitution, and then solving for the distance $d$.
Given a particular observation $\Nobst{t} = \sx{}$ (and not the expected
observation $\Nobsavgtx{\A}$) at a particular time, it can be shown that
(\ref{trans1_impulse}) re-arranges as
\begin{equation}
\label{gen_distance}
d = \vpara t \pm
\sqrt{4\Dx{}t\ln\left(\frac{\Nx{A_{EM}}V_{RX}}{\sx{}(4\pi\Dx{}t)^{3/2}}\right)
-t^2(\vperp^2+4\kth{}\Dx{})},
\end{equation}
and this equation still applies in the absence of flow and molecule
degradation, i.e., if $\vpara = \vperp = 0$ and $\kth{} = 0$. We note that the
``$\pm$'' in (\ref{gen_distance}) means that there could be two valid solutions
for $d$ if $\vpara > 0$. At any time $t$, the largest number of molecules along
the $\x$-axis is expected at the point $\{\vpara t-d,0,0\}$, and the
distribution of molecules expected along the $\x$-axis is symmetric about that
point. The ``$\pm$'' in (\ref{gen_distance}) represents uncertainty by the RX
about whether $d > \vpara t$ or $d < \vpara t$.
In this paper, if we evaluate (\ref{gen_distance}) and find two valid solutions, then we choose one via an unbiased coin toss.

There are three cases where (\ref{gen_distance}) could result in no valid
distance, as follows:
\begin{enumerate}
	\item If the observation $\sx{} = 0$, then the solution is $d = \infty$. This is more likely if the observation time is long before or long after the peak of the channel impulse response.
	\item If $\sx{}$ is sufficiently large relative to the other variables in the logarithm, then the logarithm can evaluate to a negative value and $d$ would then be a complex number. This is more likely to occur if the observation is made when a large number of molecules is expected.
	\item $d$ could be negative if $\vpara < 0$.
\end{enumerate}

In this paper, we deal with the first case by setting $\sx{} = 0.1$ and then solving for $d$. We deal with the second and third cases by setting $d = 0$. Alternative strategies can be considered in future work.

Some existing distance estimation protocols require detecting
when the peak number of molecules is observed.
By taking the derivative of (\ref{trans1_impulse}) with respect to $t$ and
setting it equal to 0, it can be shown that the peak number of
molecules at the RX, due to an instantaneous release of $\A$ molecules by
the TX at time $t=0$, would be expected at time
\begin{equation}
\label{tmax_gen}
t_{max} = \left(-3+\sqrt{9+d^2\eta/\Dx{}}\right)/\eta,
\end{equation}
where
\begin{equation}
\label{def_eta}
\eta = (\vpara^2 + \vperp^2)/\Dx{} + 4\kth{}.
\end{equation}

Interestingly, (\ref{tmax_gen}) shows that the \emph{direction} of flow has
\emph{no impact} on the time when the peak number of molecules is expected; only
the magnitude of the flow matters. Thus, (\ref{tmax_gen}) is also the time when
the peak number of molecules would be expected at the TX due
to an instantaneous release of molecules by
the RX at time $t=0$.

In the absence of flow and molecule
degradation, i.e., if $\vpara = \vperp = 0$ and $\kth{} = 0$, then it can be
shown that the peak number of molecules would be expected at the RX
at time
\begin{equation}
\label{tmax_simple}
t_{max} = d^2/(6\Dx{}).
\end{equation}

If we substitute (\ref{tmax_simple}) into (\ref{trans1_impulse}), then we can
write the number of molecules expected to be observed at $t_{max}$ in the
absence of flow and molecule degradation as
\begin{equation}
\Nobsavg{t_{max}} =
\frac{\Nx{A_{EM}}V_{RX}}{(2\pi/3)^{3/2}d^3}\EXP{-\frac{3}{2}}.
\end{equation}

\subsection{Existing Distance Estimation Protocols}

The existing distance estimation protocols that we have selected were generally
chosen for their accuracy and all rely on impulses sent by the TX at time
$t = 0$. In order to maintain a consistent comparison between protocols, and
also to facilitate tractable analysis, we consider all protocols in one-way
form. We transform two-way protocols into one-way protocols with the
understanding that we modify the knowledge required to implement those
protocols.
However, we note that, given the required knowledge, the one-way form of a
protocol should be no less accurate than its two-way form. This is because a
one-way method estimates the distance from one impulse signal. A two-way method
uses a cascade of two impulse signals where the release of the second
impulse depends on the detection of the first.

We describe the selected protocols in one-way form as follows:
\begin{itemize}
  \item The round-trip time from threshold concentration (RTT-T) protocol was
  proposed in \cite{RefWorks:614} as a two-way method without synchronization.
  In one-way form, the RX is synchronized with the TX and has a pre-determined
  threshold observation $\tau$.
  When the RX observes a number of $\A$ molecules that is
  greater than or equal to $\tau$, it substitutes the current time $t$ into
  (\ref{gen_distance}), sets $\sx{}=\tau$, and solves for $d$.
  \item The signal attenuation with timer (SA-T) protocol is another two-way
  protocol that was proposed in \cite{RefWorks:614} but was shown to be a generally inaccurate protocol. We consider it here because we will
  show in Section~\ref{sec_est} that its performance is effectively equivalent
  to the maximum likelihood distance estimate for one observation.
  In one-way form, the RX has a
  pre-determined observation time $t_{SA}$, when the current observation
  $\sx{}$ is substituted into (\ref{gen_distance}) and the RX solves for $d$.
  \item The envelope detector (ENVD) protocol was proposed in
  \cite{RefWorks:776} as a one-way protocol without
  synchronization. The RX tries to estimate the expected
  peak $\A$ molecule concentration by tracking the upper and lower envelopes of
  the observations. It is assumed that the time-varying mean of the two
  envelopes represents the true expected molecule concentration. The
  peak value of the mean of the two envelopes, $\tilde{\sx{}}$, is substituted
  for $\sx{}$ in (\ref{gen_distance}), and $t$ is replaced with either
  (\ref{tmax_gen}) or (\ref{tmax_simple}) as appropriate, so that the RX can
  solve for $d$. In the absence of flow and molecule degradation, substituting
  (\ref{tmax_simple}) into (\ref{gen_distance}) enables us to solve for $d$
  explicitly, such that we can write the estimate as
  \begin{equation}
  \hat{d}\textBar{ENVD} =
  \left(2\pi e/3\right)^{-\frac{1}{2}}
  \sqrt[3]{\Nx{A_{EM}}V_{RX}/\tilde{\sx{}}}.
  \end{equation}
  
  In the presence of flow or molecule degradation, we must solve
  (\ref{gen_distance}) for $d$ numerically because of the $d^2$
  term inside the square root in (\ref{tmax_gen}).
\end{itemize}

\section{Bound on Distance Estimation}
\label{sec_bound}

In this section, we derive the Cramer-Rao lower bound (CRLB) on the variance of
any \emph{unbiased} distance estimation protocol. Due to the noise of diffusion,
some of the protocols studied here are biased, i.e., the expected values of
their estimates are not equal to the true distance. Nevertheless,
the CRLB will provide insight for comparing protocols. We derive the CRLB for an
arbitrary number of samples, $\M$, taken by the RX.

To derive the CRLB, we need the joint probability distribution function
(PDF) of the RX's $\M$ observations. The TX releases $\A$ molecules at time $t =
0$ and then the $\smM$th observation, $\sx{\smM}$, is taken at time $t_{\smM}$.
The vector $\vec{\sx{}} = [\sx{1},\ldots,\sx{\M}]$ contains all $\M$
observations. We assume that the time between consecutive observations is
sufficient for each observation $\sx{\smM}$ to be independent; see
\cite{RefWorks:747} for a detailed discussion of observation independence
(strictly speaking, protocols that are designed to make observations
continuously, such as the RTT-T protocol,
should be sampling so fast that consecutive observations
cannot be independent). Furthermore, we will assume that the individual
observations are Poisson random variables whose means are the expected values
of the observations at the corresponding times (this has been shown to be highly
accurate in our previous work, including \cite{RefWorks:752}).
Thus, the joint PDF of the RX's observations, $p(\vec{\sx{}};d)$, is
\begin{equation}
\label{joint_pdf}
p(\vec{\sx{}};d) = \prod_{\smM = 1}^\M
{\Nxtavg{t_{\smM}}{obs}}^{\sx{\smM}}\EXP{-\Nxtavg{t_{\smM}}{obs}}/\sx{\smM}!,
\end{equation}
where $\Nxtavg{t_{\smM}}{obs}$ is as given by (\ref{trans1_impulse}). We
write
\begin{equation}
\Nxtavg{t_{\smM}}{obs} = \constAx{\smM}\EXP{-d^2\constBx{\smM} + d\constCx{}},
\end{equation}
for compactness, where
\begin{align}
\constAx{\smM} = & \;\frac{\Nx{A_{EM}}V_{RX}}{(4\pi \Dx{}
t_{\smM})^{3/2}}
\EXP{-\kth{}t_{\smM} -
\frac{t_{\smM}}{4\Dx{}}\left(\vpara^2+\vperp^2\right)}, \nonumber \\
\constBx{\smM} = & \;1/\left(4\Dx{}t_{\smM}\right), \qquad \constCx{} =
\vpara/\left(2\Dx{}\right).
\end{align}

For the CRLB to exist, the regularity condition must be satisfied, i.e., for
all $d$,
\cite[Ch. 3]{RefWorks:803}
\begin{equation}
\label{regularity}
E\left[\pbypx{\ln p(\vec{\sx{}};d)}{d}\right] = 0,
\end{equation}
where $E\left[\cdot\right]$ is the expectation taken with respect to
$p(\vec{\sx{}};d)$. If (\ref{regularity}) is satisfied, then the CRLB on the
variance of any unbiased estimator $\hat{d}$ is \cite[Eq. 3.7]{RefWorks:803}
\begin{equation}
\label{crlb_def}
\var{\hat{d}} \ge -E\left[\psbypxs{\ln p(\vec{\sx{}};d)}{d}\right]^{-1}.
\end{equation}

We now present the following theorem:

\begin{theorem}[CRLB for distance estimation]
\label{theorem_crlb}
The lower bound on the variance of any unbiased distance estimator, that is
evaluated using $\M$ independent observations of the channel impulse response,
is
\begin{equation}
\label{crlb}
\var{\hat{d}} \ge \frac{4{\Dx{}}^2}
{\sum_{\smM =
1}^\M\left(\vpara-\frac{d}{t_{\smM}}\right)^2\Nxtavg{t_{\smM}}{obs}}.
\end{equation}
\end{theorem}
\begin{IEEEproof}
Using the properties of logarithms and exponentials, it can be shown that
\begin{align}
\pbypx{\ln p(\vec{\sx{}};d)}{d} = &\; \sum_{\smM =1}^\M
\big[\!-2\sx{\smM}\constBx{\smM}d + \sx{\smM}\constCx{}
-\constAx{\smM}(\constCx{}-2\constBx{\smM}d) \nonumber \\
& \;\times\EXP{-d^2\constBx{\smM}
+ d\constCx{}}\!\big],
\end{align}
and, by recalling that $E\left[\sx{\smM}\right] = \Nxtavg{t_{\smM}}{obs}$, we
can conclude that the regularity condition (\ref{regularity}) is satisfied.

It can then be shown that
\begin{align}
E\left[\psbypxs{\ln p(\vec{\sx{}};d)}{d}\right] = &\;
- \sum_{\smM =1}^\M \constAx{\smM}(\constCx{}-2\constBx{\smM}d)^2 \nonumber \\
& \;\times\EXP{-d^2\constBx{\smM}+d\constCx{}},
\end{align}
which we substitute into (\ref{crlb_def}). We substitute the values of
$\constAx{\smM}$, $\constBx{\smM}$, and $\constCx{}$ to arrive at (\ref{crlb}).
\end{IEEEproof}

We note that if we had assumed that the observations follow Gaussian instead
of Poisson statistics, then the regularity condition (\ref{regularity}) would
not be satisfied and the CRLB would therefore not exist.

Eq. (\ref{crlb}) gives us insight into the factors affecting the
accuracy of a distance estimate. A more accurate estimate might be
possible if more samples are taken (i.e., by increasing $\M$). Also, the
variance should decrease if the expected values of the samples increase, e.g.,
by the TX releasing more $\A$ molecules or
by decreasing molecule degradation rate $\kth{}$. These inferences
might be intuitive, but the derivation of (\ref{crlb}) gives them theoretical
justification. The impact of some parameters, such as the diffusion coefficient
$\Dx{}$, are less obvious because they are both inside and outside the
$\Nxtavg{t_{\smM}}{obs}$ term in (\ref{crlb}).

In our simulations, we compare the CRLB with an estimator's mean square
error, because the mean square error enables direct comparisons between biased
and unbiased estimators. The mean square estimation error and the
variance of the estimation error, given as
\begin{equation}
\label{mse_vs_var}
\mse{\hat{d}} = E\!\left[\!\left(\hat{d} - d\right)^2\!\right], \quad
\var{\hat{d}} = E\!\left[\!\left(\hat{d} - E\Big[\hat{d}\Big]\right)^2\!\right],
\end{equation}
respectively, are equivalent only if the estimator
is unbiased, i.e., if $E\Big[\hat{d}\Big] = d$, as discussed in \cite[Ch.
2]{RefWorks:803}.
The mean square error of a biased estimator such as the ENVD protocol is
composed of errors due to the estimator's variance in addition to its bias.

\section{Maximum Likelihood Estimation}
\label{sec_est}

In this section, we derive the maximum likelihood (ML) distance estimate of a
one-way protocol for any number of samples taken by the RX. Using
one sample is a special case that is shown to be equivalent to the SA-T
protocol.

The ML distance estimate $\hat{d}$ is the distance that
maximizes the joint observation likelihood $p(\vec{\sx{}};d)$. We try to find
this distance by taking the partial derivative of (\ref{joint_pdf}) with
respect to $d$ and setting it equal to 0. It can be shown that finding $d$ such
that $\pbypx{p(\vec{\sx{}};d)}{d} = 0$, under meaningful physical parameters
(i.e., finite RX volume $V_{RX}$, finite distance $d$, etc.), is
equivalent to finding $d$ such that $\pbypx{\ln p(\vec{\sx{}};d)}{d} = 0$. This
is the same as satisfying the regularity condition in (\ref{regularity}) but
\emph{without} the expectation. In other words, the ML estimate
is the distance that satisfies
\begin{multline}
\label{ml_est}
\sum_{\smM =1}^\M
\bigg[\!-2\sx{\smM}\constBx{\smM}\hat{d} + \sx{\smM}\constCx{} \\
-\constAx{\smM}(\constCx{}-2\constBx{\smM}\hat{d})
\EXP{-{\hat{d}}^2\constBx{\smM} + \hat{d}\constCx{}}\!\bigg] = 0.
\end{multline}

If $\M = 1$, then it can be shown that (\ref{ml_est}) is satisfied if the
observation $\sx{1} = \constAx{1}\EXP{-{\hat{d}}^2\constBx{1} +
\hat{d}\constCx{}}$, i.e., $\hat{d}$ must be the distance whose expected
observation $\Nxtavg{t_{1}}{obs}$ is equal to $\sx{1}$. Thus, $\hat{d}$ can be
found by substituting $\sx{1}$ and $t_1$ into (\ref{gen_distance}).

We emphasize that the ML estimate for a \emph{given} time $t_1$ uses
the observation made at that time to estimate $d$. Hence, the ML
estimate for $\M=1$ is effectively equivalent to the SA-T protocol, even though
this protocol was shown to have poor performance in 
\cite{RefWorks:614}. The reason for its poor performance is that all other
existing protocols track the observed signal over time until some
criterion is met, i.e., information is combined from multiple observations in
order to measure the distance (even though the final calculation might only use
the value of a single observation). The SA-T does \emph{not} track the signal
over time but makes the ML estimate for its one observation.

If $\M > 1$, then a solution to (\ref{ml_est}) is non-trivial.
We approximate the
true ML distance by performing a discretized 1-dimensional search (over non-negative $d$) and choose the distance that maximizes the log likelihood $\ln p(\vec{\sx{}};d)$ given
the vector $\vec{\sx{}}$ of $\M$ observations. We impose a finite upper bound on $d$, so the estimate is always finite.

For any value of $\M$, the RX
must know the values of all other environmental parameters, i.e., $\vpara$,
$\vperp$, $\Dx{}$, $\Nx{A_{EM}}$, $V_{RX}$, and $\kth{}$, and be synchronized
with the TX, in order to make a ML estimate. This is intuitive; an ML estimate
must make use of all knowledge available. Even though the RTT-T and ENVD
protocols also require knowledge of all other environmental parameters (in the
general case with flow and molecule degradation), they cannot be ML estimates.
They make multiple observations but only use a single observation
to estimate the distance (actually, in the case of the ENVD protocol, the mean
of two filtered observations is used), discarding the information
available from all other observations.

\section{Numerical Results}
\label{sec_results}

In this section, we present simulation results to assess the performance
of the distance estimation protocols discussed in this paper, particularly with
respect to the CRLB. Our simulations were executed in the particle-based
stochastic framework that we described in \cite{RefWorks:662}. Every molecule
released by the TX is treated as an independent particle whose location is
updated every simulation time step $\Delta t$. The probability of a given
molecule degrading via reaction (\ref{k_mechanism}) in one time step is
$\kth{}\Delta t$. We consider two sets of environmental parameters as listed in
Table~\ref{table_param}. The chosen values are consistent with those that we
have considered in our previous work, including \cite{RefWorks:752}.
The molecule degradation rate for System 2,
$\kth{}=62.5\,\second^{-1}$, was chosen so that, in the absence of flow, one
less molecule is expected at the expected peak concentration time than if
$\kth{} = 0$, i.e., $\Nobsavg{t_{max}} = 6.5$ instead of $7.5$.
We note that we set the velocity of flow perpendicular to the line between the
TX and RX (i.e., $\vperp$) to 0. Simulations are averaged over $10^4$
independent realizations.

\begin{table}[!tb]
	\centering
	\caption{System parameters used for numerical and simulation results.}
	{\renewcommand{\arraystretch}{1.4}
		\begin{tabular}{|l|c|c|c|}
		\hline
		Parameter & Symbol & System 1 & System 2\\ \hline
		Distance Between TX and RX & $d$ & Various & $4\,\mu\metre$	\\ \hline
		Flow From TX to RX & $\vpara$ & 0 & Various	\\ \hline
		Degradation Rate & $\kth{}$ & 0 & $62.5\,\second^{-1}$	\\ \hline
		Molecules per TX emission & $\Nx{A_{EM}}$ & \multicolumn{2}{c|}{$10^5$}	\\ \hline
		Diffusion coefficient
		& $\Dx{\A}$ & \multicolumn{2}{c|}{$10^{-9}\,\metre^2/\second$}
		 \\ \hline
		Radius of RX & $r_{RX}$	& \multicolumn{2}{c|}{$0.5\,\mu\metre$}  \\ \hline
		Simulation Time Step & $\Delta t$ & \multicolumn{2}{c|}{$0.1\,\metre\second$}
		 \\ \hline
		Number of Simulation Steps & - & \multicolumn{2}{c|}{$200$}
		 \\ \hline
		\end{tabular}
	}
	\label{table_param}
\end{table}

\begin{figure}[!tb]
\centering
\includegraphics[width=\linewidth]{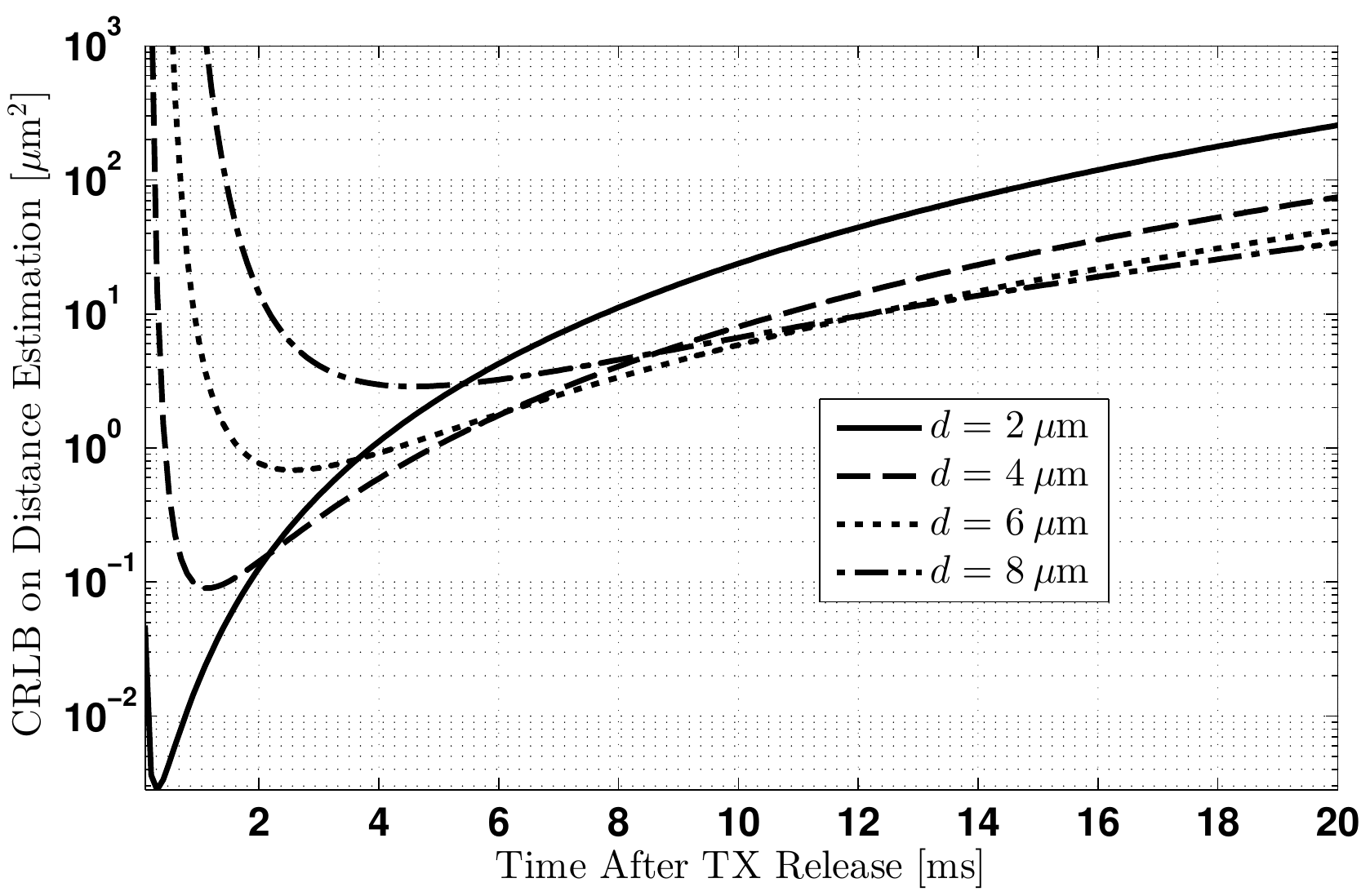}
\caption{The CRLB on the variance of the error in distance estimation in
System 1 as a function of time at varying distances from the TX.}
\label{fig_crlb}
\end{figure}

Prior work on distance estimation has been applied in the absence of
flow and molecule degradation, i.e., when $\vpara = \vperp = 0$ and $\kth{} =
0$, so most of the results that we present here focus on a similar environment,
i.e., System 1, but in 3 dimensions.
In Fig.~\ref{fig_crlb}, we show how the CRLB varies in System 1 over distance
and time for a single observation, i.e., for $\M = 1$. An estimate made too
soon will have poor accuracy at any distance, because no molecules would have
arrived at the RX. An observation made after a long time will also result in
poor estimation accuracy because most of the molecules released would have
diffused away.
Intuitively, the best time for an estimate is during the fastest change in the
expected number of molecules. This is confirmed by Fig.~\ref{fig_crlb}, where
the minimum variance occurs approximately halfway between when the TX releases
the $\A$ molecules and when the peak number of molecules is expected at the RX.
For example, we can calculate from (\ref{tmax_simple}) that when
$d=4\,\mu\metre$, $\tmax = 2.66\,\metre\second$, and Fig.~\ref{fig_crlb} shows
that the CRLB's lowest value at that distance is about $1.2\,\metre\second$
after release by the TX.
At shorter distances, the CRLB is much more sensitive over time but its minimum
is much lower, due to the sudden arrival of a large number of
molecules followed by a relatively more rapid dissipation. This can be observed
for $d=2\,\mu\metre$, where the minimum CRLB is more than an order of magnitude
lower than for $d=4\,\mu\metre$ but becomes higher when time $t >
2.5\,\metre\second$.

\begin{figure}[!tb]
\centering
\includegraphics[width=\linewidth]{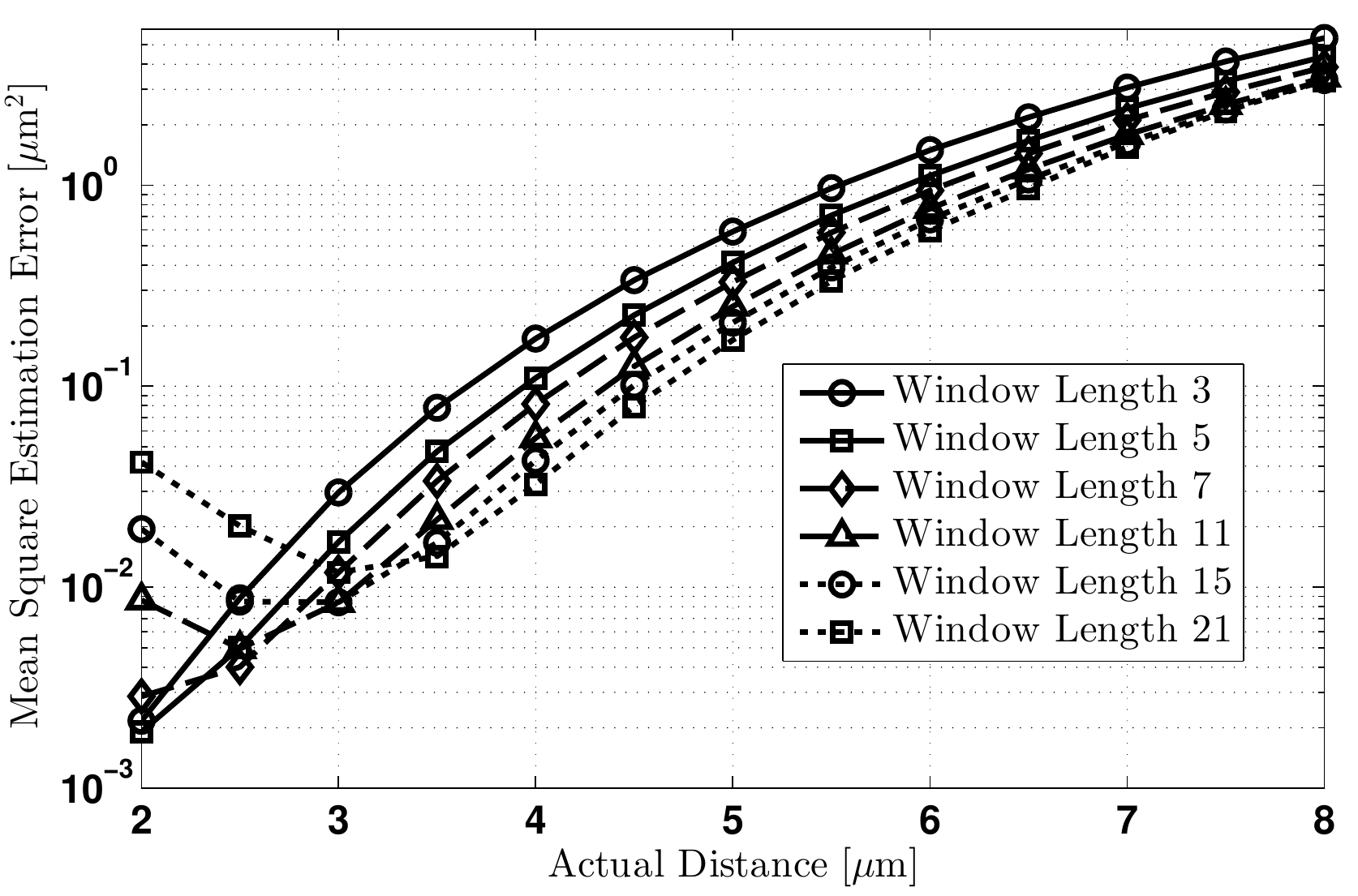}
\caption{Mean square error of the ENVD protocol in System 1 for varying filter
window length as a function of true distance $d$.}
\label{fig_envd}
\end{figure}

When the ENVD protocol was introduced in \cite{RefWorks:776}, the authors did
not describe their implementation of the upper and lower envelope detectors. We
implemented what we refer to as a \emph{moving maximum filter} and a
\emph{moving minimum filter}, where each filtered observation is found by taking
the maximum and minimum of the nearest (in time) RX observations for the upper
and lower envelopes, respectively.
In Fig.~\ref{fig_envd}, we evaluate the mean square estimation error of the ENVD
protocol in System 1 for varying filter window length as a function of the true
distance $d$. Shorter filter lengths are better at shorter distances because the
diffusion wave rises and falls more abruptly when the RX is closer to
TX. At longer distances, the rise and fall of the diffusion wave is more gradual
and so longer filter lengths are more accurate.
For example, the shortest window length, 3, is the second best for $d =
2\,\mu\metre$ and the worst for $d \ge 3\,\mu\metre$. We choose filter window
length 7 for comparison with the other distance estimation protocols.

\begin{figure}[!tb]
\centering
\includegraphics[width=\linewidth]{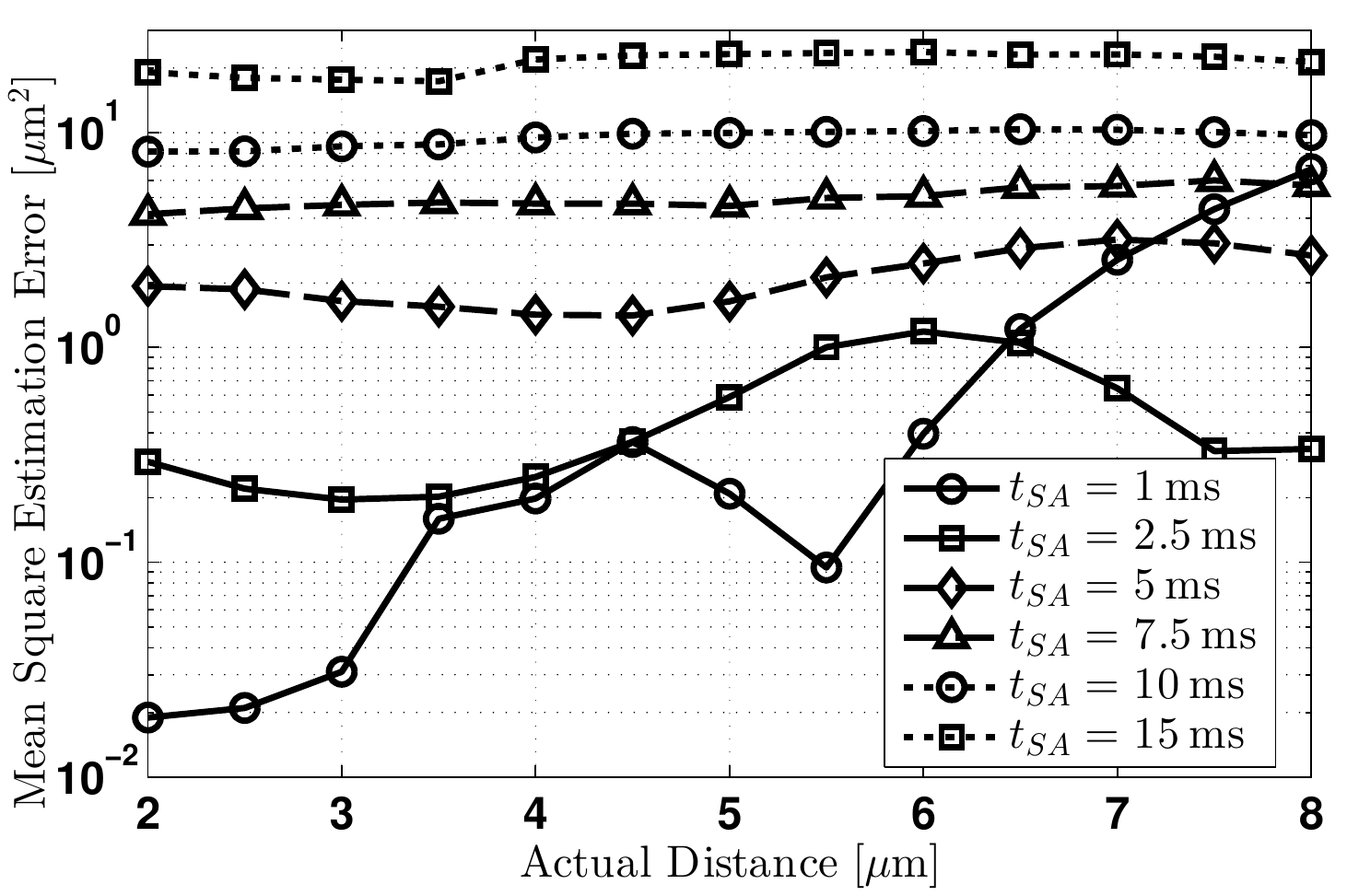}
\caption{Mean square error of the SA-T
protocol in System 1 for varying observation time $t_{SA}$ as a function of
true distance $d$.}
\label{fig_sat}
\end{figure}

In Fig.~\ref{fig_sat}, we evaluate the mean square estimation error of the SA-T
protocol in System 1 for varying sample time $t_{SA}$ as a function of the true
distance $d$.
We observe that earlier sampling times are generally better, although sampling at time $t_{SA} = 2.5\,\metre\second$ has more consistent performance over the range of $d$ than sampling at $t_{SA} = 1\,\metre\second$. The sampling time $t_{SA} = 2.5\,\metre\second$ has fewer occurrences of observations that would, without the corrections proposed in Section~\ref{sec_model}, result in no valid estimate. Due to lack of space, we do not present detailed results of how frequently corrections are required. We choose sampling time $t_{SA} = 2.5\,\metre\second$ for comparison with the other protocols in the remaining figures.

\begin{figure}[!tb]
\centering
\includegraphics[width=\linewidth]{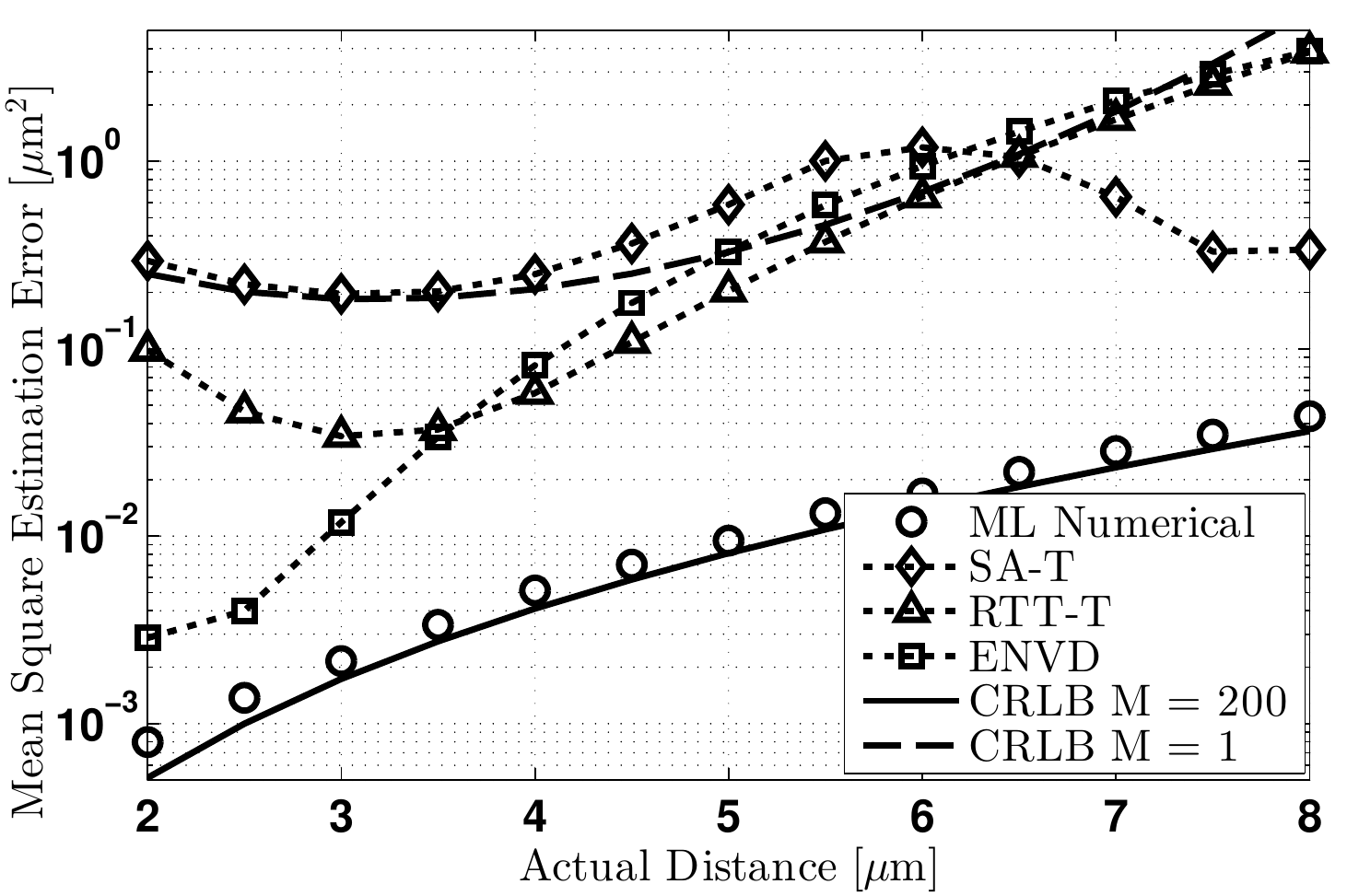}
\caption{Mean square error in System 1, of all
distance estimation protocols being considered,
as a function of the true distance $d$. The CRLB is also shown for
$\M = 1$ (to compare with the SA-T protocol) and for $\M = 200$ (to compare
with all other protocols).}
\label{fig_all}
\end{figure}

In Fig.~\ref{fig_all}, we compare the mean square error in System 1, of
all distance estimation protocols being considered,
as a function of the true distance $d$. For the RTT-T protocol, we chose
threshold $\tau=2$ for its overall performance over the entire range of $d$
(a figure similar to Fig.~\ref{fig_sat} but for the RTT-T protocol and
varying $\tau$ instead of $t_{SA}$, is not shown).
We compare the SA-T protocol with the CRLB for $\M=1$ and evaluated at the
time $t = 2.5\,\metre\second$. The ``numerical'' ML curve was found by
calculating the log likelihood $\ln p(\vec{\sx{}};d)$ over a range of distances
from $0.01\,\mu\metre$ to $20\,\mu\metre$, given \emph{all} 200 samples in
$\vec{\sx{}}$, and selecting the distance with
the largest log likelihood. The CRLB for $\M = 200$ was evaluated by solving
(\ref{crlb}) for all 200 samples. For clarity of exposition, we do not
evaluate the CRLB specifically for the ENVD and RTT-T protocols; the number of
samples $\M$ that they observe for a single distance estimate can change with
every realization.

Fig.~\ref{fig_all} shows that the numerical ML estimate is a lower bound on all
distance estimation protocols, and it is also very close to achieving the CRLB
given all observations made at the RX
(the ML estimate can move closer to the CRLB by decreasing $\M$ so that
the observations become more independent and by increasing the resolution
of the search over $d$).
Thus, the ML estimate is effectively the optimal estimate for $\M=200$.
The accuracy of the SA-T protocol is quite poor. However, at shorter
distances, i.e., $d \le 4\,\mu\metre$, it rarely requires sample corrections and its accuracy is quite
close to the CRLB when $\M=1$.
For $d \le 4\,\mu\metre$, it is practically the most accurate unbiased estimate possible for a single observation, and it is equivalent to the single-sample ML estimate when sample corrections are not required.
At higher distances, i.e., $d \ge 7\,\mu\metre$, the SA-T protocol appears
to be \emph{more} accurate than the CRLB for $\M=1$, and this is because
frequent observations of $\sx{1}=0$ that are corrected to $0.1$ introduce an estimate bias that improves the mean square error over the unbiased case.
Finally, we observe that the RTT-T and ENVD protocols are generally more
accurate than the SA-T protocol but much less accurate than the CRLB evaluated
for $\M=200$.

\begin{figure}[!tb]
\centering
\includegraphics[width=\linewidth]{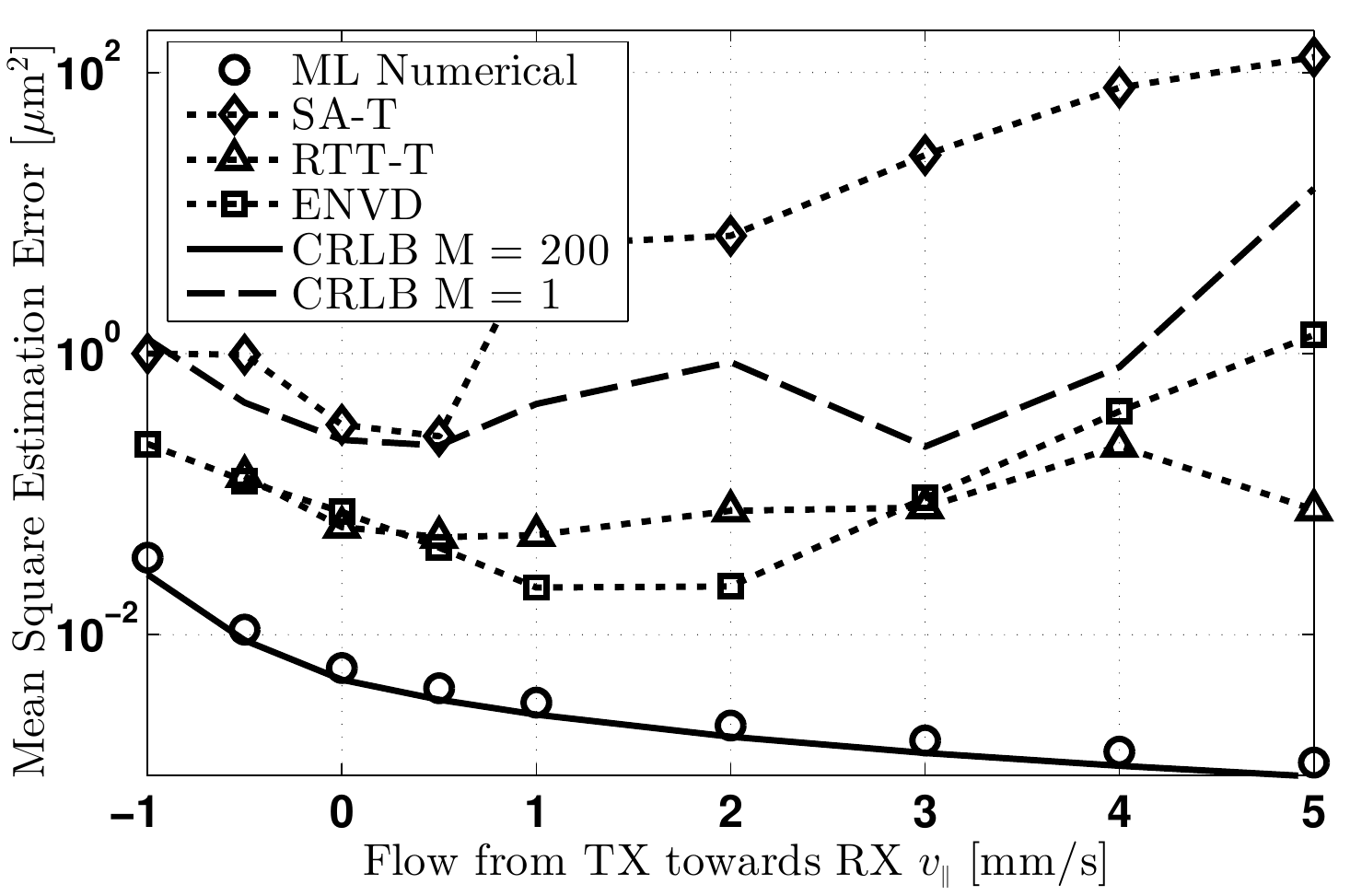}
\caption{Mean square error in System 2,
of all distance estimation protocols being considered,
as a function of the flow $\vpara$. The CRLB is also
shown for $\M = 1$ and for $\M = 200$.}
\label{fig_all_flow}
\end{figure}

In Fig.~\ref{fig_all_flow}, we compare the mean square error in System 2, of all distance estimation protocols being
considered, as a function of the flow velocity from the TX towards the RX,
$\vpara$. We choose the single distance $d = 4\,\mu\metre$ because of the
accuracy of all protocols at that distance in Fig.~\ref{fig_all}.
All protocols have the same configuration as described for System 1.
We see that
the numerical ML estimate is still a lower bound on all distance estimation
protocols, and it is still very close to the CRLB for $\M = 200$.
The accuracy of the SA-T protocol is also still close to the CRLB for $\M=1$
for low values of $\vpara$, but degrades for $\vpara \ge 1\,\metre\metre/\second$ because of the increasing frequency of corrected estimates that introduce a destructive bias and because we often need to choose between two valid distances via a coin toss (due to the ``$\pm$'' in (\ref{gen_distance})).

\section{Conclusions}
\label{sec_concl}

In this paper, we studied distance estimation for diffusive molecular
communication in an idealized 3-dimensional environment. We derived the CRLB on the error variance of any unbiased
distance estimator at the RX taking independent samples of an impulsive signal
from the TX. The CRLB is a function of the samples expected at the RX. We
derived the ML estimator and showed that, in the single-sample
case, it is equivalent to the SA-T protocol.
A numerical evaluation of the ML estimator that uses all samples at the RX was
shown to achieve the CRLB and be more accurate than a selection of the
most accurate existing distance estimation protocols.

Our approach for bounding the accuracy of distance estimation protocols can
serve as a benchmark for the design of future protocols and as a guide for
finding the optimal estimate of other diffusive molecular communication
parameters as well as the channel impulse response. Ultimately, we are
interested in the design of low-complexity estimators for more realistic environments. Low-complexity protocols might be
more practical, and bounds on accuracy give us insight into how
much we lose by implementing sub-optimal protocols.

\bibliography{../references/nano_ref}

\begin{thebibliography}{10}
\providecommand{\url}[1]{#1}
\csname url@samestyle\endcsname
\providecommand{\newblock}{\relax}
\providecommand{\bibinfo}[2]{#2}
\providecommand{\BIBentrySTDinterwordspacing}{\spaceskip=0pt\relax}
\providecommand{\BIBentryALTinterwordstretchfactor}{4}
\providecommand{\BIBentryALTinterwordspacing}{\spaceskip=\fontdimen2\font plus
\BIBentryALTinterwordstretchfactor\fontdimen3\font minus
  \fontdimen4\font\relax}
\providecommand{\BIBforeignlanguage}[2]{{%
\expandafter\ifx\csname l@#1\endcsname\relax
\typeout{** WARNING: IEEEtran.bst: No hyphenation pattern has been}%
\typeout{** loaded for the language `#1'. Using the pattern for}%
\typeout{** the default language instead.}%
\else
\language=\csname l@#1\endcsname
\fi
#2}}
\providecommand{\BIBdecl}{\relax}
\BIBdecl

\bibitem{RefWorks:801}
T.~Nakano, A.~Eckford, and T.~Haraguchi, \emph{Molecular Communication}.\hskip
  1em plus 0.5em minus 0.4em\relax Cambridge University Press, 2013.

\bibitem{RefWorks:588}
B.~Alberts, D.~Bray, K.~Hopkin, A.~Johnson, J.~Lewis, M.~Raff, K.~Roberts, and
  P.~Walter, \emph{Essential Cell Biology}, 3rd~ed.\hskip 1em plus 0.5em minus
  0.4em\relax Garland Science, 2010.

\bibitem{RefWorks:644}
H.~ShahMohammadian, G.~G. Messier, and S.~Magierowski, ``Optimum receiver for
  molecule shift keying modulation in diffusion-based molecular communication
  channels,'' \emph{Nano Commun. Net.}, vol.~3, no.~3, pp. 183--195, Sep. 2012.

\bibitem{RefWorks:747}
A.~Noel, K.~C. Cheung, and R.~Schober, ``Optimal receiver design for diffusive
  molecular communication with flow and additive noise,'' \emph{IEEE Trans.
  Nanobiosci.}, vol.~13, no.~3, pp. 350--362, Sep. 2014.

\bibitem{RefWorks:786}
D.~Kilinc and O.~B. Akan, ``Receiver design for molecular communication,''
  \emph{IEEE J. Sel. Areas Commun.}, vol.~31, no.~12, pp. 705--714, Dec. 2013.

\bibitem{RefWorks:608}
T.~Nakano, M.~J. Moore, F.~Wei, A.~V. Vasilakos, and J.~Shuai, ``Molecular
  communication and networking: Opportunities and challenges,'' \emph{IEEE
  Trans. Nanobiosci.}, vol.~11, no.~2, pp. 135--148, Jun. 2012.

\bibitem{RefWorks:488}
M.~J. Moore and T.~Nakano, ``Addressing by beacon distances using molecular
  communication,'' \emph{Nano Commun. Net.}, vol.~2, no. 2-3, pp. 161--173,
  Jun. 2011.

\bibitem{RefWorks:802}
M.~J. Moore, T.~Nakano, A.~Enomoto, and T.~Suda, ``Measuring distance with
  molecular communication feedback protocols,'' in \emph{Proc. ICST BIONETICS},
  Dec. 2010, pp. 1--13.

\bibitem{RefWorks:614}
------, ``Measuring distance from single spike feedback signals in molecular
  communication,'' \emph{IEEE Trans. Signal Process.}, vol.~60, no.~7, pp.
  3576--3587, Jul. 2012.

\bibitem{RefWorks:776}
J.~T. Huang, H.~Y. Lai, Y.~C. Lee, C.~H. Lee, and P.~C. Yeh, ``Distance
  estimation in concentration-based molecular communications,'' in \emph{Proc.
  IEEE GLOBECOM}, Dec. 2013, pp. 2587--2597.

\bibitem{RefWorks:674}
M.~J. Moore and T.~Nakano, ``Comparing transmission, propagation, and receiving
  options for nanomachines to measure distance by molecular communication,'' in
  \emph{Proc. IEEE ICC}, Jun. 2012, pp. 6132--6136.

\bibitem{RefWorks:761}
H.~ShahMohammadian, G.~Messier, and S.~Magierowski, ``Blind synchronization in
  diffusion-based molecular communication channels,'' \emph{IEEE Commun.
  Letters}, vol.~17, no.~11, pp. 2156--2159, Nov. 2013.

\bibitem{RefWorks:773}
M.~J. Moore and T.~Nakano, ``Oscillation and synchronization of molecular
  machines by the diffusion of inhibitory molecules,'' \emph{IEEE Trans.
  Nanotechnol.}, vol.~12, no.~4, pp. 601--608, Jul. 2013.

\bibitem{RefWorks:752}
A.~Noel, K.~C. Cheung, and R.~Schober, ``Diffusive molecular communication with
  disruptive flows,'' in \emph{Proc. IEEE ICC}, Jun. 2014, pp. 3600--3606.

\bibitem{RefWorks:803}
S.~M. Kay, \emph{Fundamentals of Statistical Signal Processing: Estimation
  Theory}, 1993, vol.~1.

\bibitem{RefWorks:662}
A.~Noel, K.~C. Cheung, and R.~Schober, ``Improving receiver performance of
  diffusive molecular communication with enzymes,'' \emph{IEEE Trans.
  Nanobiosci.}, vol.~13, no.~1, pp. 31--43, Mar. 2014.

\end{thebibliography}

\end{document}